\documentclass[letterpaper]{article} 
\usepackage[submission]{aaai24}  
\usepackage{times}  
\usepackage{helvet}  
\usepackage{courier}  
\usepackage[hyphens]{url}  
\usepackage{graphicx} 
\usepackage{natbib}  
\usepackage{caption} 
\usepackage{algorithm}
\usepackage{algorithmic}
\usepackage{cleveref}
\usepackage{quoting}
\quotingsetup{font={itshape,footnotesize},vskip=2pt, leftmargin=15pt}

\usepackage{newfloat}
\usepackage{listings}
\DeclareCaptionStyle{ruled}{labelfont=normalfont,labelsep=colon,strut=off} 
\floatstyle{ruled}
\newfloat{listing}{tb}{lst}{}
\floatname{listing}{Listing}
\title{Unreflected Acceptance - Investigating the Negative Consequences of ChatGPT-Assisted Problem Solving in Physics Education}

\newcommand{\samplemean}{\bar{x}}
\newcommand{\samplesd}{s}
\newcommand{\baselinecondition}{\textsc{Search Engine}}
\newcommand{\chatgptcondition}{\textsc{ChatGPT}}

\author {
    Lars Krupp\textsuperscript{\rm 1, \rm 2},
    Steffen Steinert\textsuperscript{\rm 3},
    Maximilian Kiefer-Emmanouilidis\textsuperscript{\rm 1, \rm 2},
    Karina E. Avila \textsuperscript{\rm 2},
    Paul Lukowicz \textsuperscript{\rm 1, \rm 2},
    Jochen Kuhn \textsuperscript{\rm 3},
    Stefan Küchemann \textsuperscript{\rm 3},
    Jakob Karolus \textsuperscript{\rm 1, \rm 2}
}
\affiliations {
    \textsuperscript{\rm 1}German Research Center for Artificial Intelligence (DFKI), Kaiserslautern, Germany\\
    \textsuperscript{\rm 2}RPTU Kaiserslautern-Landau, Kaiserslautern, Germany\\
    \textsuperscript{\rm 3}LMU Munich, Munich, Germany\\
}

\begin{document}

\maketitle

\begin{abstract}
Large language models (LLMs) have recently gained popularity. However, the impact of their general availability through ChatGPT on sensitive areas of everyday life, such as education, remains unclear. Nevertheless, the societal impact on established educational methods is already being experienced by both students and educators. 
Our work focuses on higher physics education and examines problem solving strategies. In a study, students with a background in physics were assigned to solve physics exercises, with one group having access to an internet search engine (N=12) and the other group being allowed to use ChatGPT (N=27). We evaluated their performance, strategies, and interaction with the provided tools. 
Our results showed that nearly half of the solutions provided with the support of ChatGPT were mistakenly assumed to be correct by the students, indicating that they overly trusted ChatGPT even in their field of expertise. 
Likewise, in 42\% of cases, students used copy \& paste to query ChatGPT --- an approach only used in 4\% of search engine queries --- highlighting the stark differences in interaction behavior between the groups and indicating limited reflection when using ChatGPT. In our work, we demonstrated a need to (1) guide students on how to interact with LLMs and (2) create awareness of potential shortcomings for users.

\end{abstract}

\section{Introduction}

LLMs have been omnipresent in media and the public eye since November 2022 when ChatGPT was first presented~\cite{openaichatgpt2022}. With one of the fastest growing user bases ever measured for any application~\cite{reuters2023a,reuters2023b}, it is difficult to estimate its future impact on every aspect of our daily lives. 

Especially in sensitive areas, such as education, easily accessible information --- true or false --- poses challenges for educators and students alike. Recent discussions around ChatGPT often involve its use in assignments, for homework and in the classroom.
At present, it is unclear how LLMs, such as ChatGPT, can meaningfully support students in educational contexts. Designing methods that allow for informed usage of these systems remains a challenge. Our work investigates how students interact with ChatGPT when given unrestricted access and highlights the need for moderated access.


LLMs are predictive models that predict the most probable next token based on a series of previously seen tokens they have already seen. As a result, they excel at tasks such as writing~\cite{yuan2022wordcraft}, translation~\cite{wang2023document}, and even programming~\cite{kashefi2023chatgpt}. Contrarily, disciplines that rely heavily on calculations and reasoning may prove more challenging for LLMs. This could lead to unforeseen or even negative consequences for students, like incorrect homework or learning an incorrect explanation of a concept. 


In our work, we examined the field of physics, specifically how students with a strong background in physics interact with ChatGPT to assist them in solving physics questions. 
We conducted a study with a total of 39 participants with backgrounds in science, technology, engineering and math (STEM) fields from multiple universities. One group (N=27) had unrestricted access to ChatGPT, while the other group (N=12) had access to a search engine.
Our findings indicate that participants with the \chatgptcondition{} condition overly trusted answers generated by ChatGPT. In particular, students often failed to recognize wrong answers given by ChatGPT and largely relied on a copy \& paste strategy to solve the posed physics questions. 
In contrast, participants in the \baselinecondition{} condition showed higher rates of reflection, as indicated by their sparse use of copy \& paste, favoring more thought-through solving strategies.


Our work shows that even students with advanced domain knowledge often struggled to differentiate between correct and incorrect answers given by LLMs and could not use the system effectively. There is a need for further research to design LLM-based support tools in a way that (1) creates awareness of their inherent uncertainty and (2) allows moderated use that encourages critical thinking.

\section{Related Work}
The field of language models (LM) offers a variety of possible applications in education. For example, they have been used for multiple-choice question generation~\cite{raina2022multiple} or answering~\cite{zhang2022greaselm}. However, since we have to expect students to use LMs like ChatGPT at home, there is a need to figure out how they utilize these powerful new tools unaided. 
In this section, we briefly introduce the current state of LMs, and chat-bots, how this relates to education, and specifically physics education.

\subsection{Language Models}
Recent advances in natural language processing, initiated by the introduction of the transformer architecture~\cite{vaswani2017attention}, have led to significant progress in the field of language models. The different approaches taken by GPT~\cite{radford2018improving} and BERT~\cite{devlin2018bert} models proved to be exceptionally successful. Progress has been steady, with a trend towards increasingly larger models, supported by their scaling laws~\cite{kaplan2020scaling}, which suggest that larger size generally leads to a better model. ChatGPT~\cite{openaichatgpt2022} brought the technology into the public eye, further accelerating the pace of publications and leading to the development of models such as LLaMA~\cite{touvron2023llama}, GPT-4~\cite{openai2023gpt4}, and PaLM-E~\cite{driess2023palm}. Some of which even support multi-modal inputs~\cite{driess2023palm}. Language models have shown their potential in many different areas~\cite{yuan2022wordcraft, wang2023document, kashefi2023chatgpt} and are a topic that also influences education~\cite{kasneci2023chatgpt}.

\subsection{Education}
LLMs offer great potential for advancing standard practices and research in education~\cite{kasneci2023chatgpt}. Several possible applications have been previously suggested, such as personalized learning, lesson planning, assessment and evaluation, to familiarize students with challenges and opportunities of LLMs~\cite{kasneci2023chatgpt}. Furthermore, a number of studies exist that investigate the use of chatbots based on different technologies in education~\cite{kuhail2023interacting}. The use of chatbots in education offers several advantages, such as serving as a pedagogical tool to help students with disabilities and to help different social groups to close the educational gap that may exist between them~\cite{perez2020rediscovering}. However, none of the systems examined in these works are based on a LLM despite several authors seeing great potential for LLM-based chatbots in the educational domain~\cite{rudolph2023war}. It should be noted that LLMs show some weaknesses. Until now, they lack higher-order thinking skills, and their outputs strongly depend on the data they have been trained on, sometimes leading to unreliable outputs~\cite{bitzenbauer2023chatgpt}. 

\subsection{Physics Education}
In physics education, there are conflicting reports on the ability of LLMs to solve physics tasks. On the one hand, a few studies have observed inconsistent behavior in ChatGPT's answers to physics questions~\cite{gregorcic2023chatgpt, santos2023enhancing}. These studies showed that ChatGPT often provides incorrect answers to physics questions and concluded that it is unsuitable as a physics tutor or for cheating on homework. This apparent weakness of ChatGPT in answering physics questions studies can be used as a learning experience to foster critical thinking skills of students. Bitzenbauer intended to foster critical thinking skills using these weaknesses by asking students to generate ChatGPT answers to a question and to discuss them critically, which led to an improved perceived usefulness of ChatGPT~\cite{bitzenbauer2023chatgpt}. On the other hand, other studies demonstrate the strength of ChatGPT 3.5 and 4.0 to solve conceptual multiple-choice questions in physics \cite{west2023ai,west2023advances}. ChatGPT was able to solve 28 out of 30 items of the force concept inventory correctly \cite{west2023advances}. Kieser and colleagues even found that ChatGPT 4.0 is able to mimic different students' difficulties when answering conceptual questions, which opens the opportunity for data augmentation, personalized support for students that is sensitive to different difficulties, and support for teachers during task creation \cite{kieser2023educational}. The latter opportunity was studied by K\"uchemann and colleagues in a randomized controlled trial who compared the characteristics and quality of created physics tasks by prospective physics teachers either using ChatGPT or a textbook. The authors found that students in both groups had difficulties with the specificity of tasks, i.e., all relevant information to solve the tasks are provided, and the students who used ChatGPT embedded the tasks less frequently in a real-world context \cite{kuchemann2023physics}. Moreover, the authors found that prospective physics teachers used the tasks in 76\% of the cases as provided by ChatGPT without modifying them \cite{kuchemann2023physics}. These findings point towards the affordances of using ChatGPT in education and the overreliance of students when using it. 
While these articles provide interesting findings and show that using ChatGPT for answering questions present great demands on students, the results were either not verified with real students~\cite{santos2023enhancing, gregorcic2023chatgpt} or the problem solving strategies when using ChatGPT were not studied ~\cite{bitzenbauer2023chatgpt, kuchemann2023physics}. To the best of our knowledge, it has not yet been investigated how students, depending on their prior knowledge, solve physics problems using ChatGPT when they have not received specific instructions or have been made aware of its limitations. This is the focus of this work.

\section{Methodology}
\label{sec:methodology}
The related work highlights the existing uncertainties regarding the use of LLMs in general and specifically in the context of physics education. However, to date, little work has been conducted that allows for moderated and informed usage of such models. We argue that informed usage of generative models is of utmost importance, particularly in educational areas. 

Our work contributes a first investigation into how students interact with LLMs and whether they are aware of their shortcomings. In a mixed-method evaluation conducted online and at two universities (RPTU Kaiserslautern-Landau and LMU Munich)
, we tasked students with solving given physics problems. Using a between-subject study design, we assessed students' performance and interaction strategies when having access to different support tools. 

In a baseline condition, students had access to an internet search engine (\baselinecondition). In the \chatgptcondition{} condition, students were able to freely use ChatGPT. We recorded the students' physics knowledge with a pretest (no support tools allowed) and their performance in the main test, as well as inquired about their impressions when interacting with ChatGPT through questionnaires and an exit interview (see~\Cref{fig:procedure}). Our research was guided by two main research questions:

\paragraph{\textbf{RQ1:} What is the performance of students when being allowed to use ChatGPT in comparison to the students who used a search engine?}
One main inquiry of our work focused on whether ChatGPT allowed students to perform better when solving the physics questions. We further analyzed the students' interaction protocols with both tools (search engine, ChatGPT) to investigate how effectively they used the tool. 


\paragraph{\textbf{RQ2:} What are predominant strategies when interacting with ChatGPT compared to search engines?}
On a meta-level, we were interested in what solving strategies students employed when using ChatGPT and how they differed from the ones used with search engines. From the conducted exit interviews, in combination with the students' interaction protocols, we distilled predominant strategies when interacting with either tool.

 
\subsection{Physics Question Acquisition}
\label{question_acquisition}
For our main test, we selected four physics questions. Since all questions should be solvable with school knowledge, we have chosen questions that require knowledge of kinematics, friction, rotational movements, inelastic collisions, conservation of energy and mathematical conversions. The tasks should be too complex for the tool in each group to answer them correctly directly or for the students to be able to solve them immediately on their own (using the tool is necessary). However, the solution should still be obtainable either by asking ChatGPT clever questions or composing good queries using the search engine and solving the problem step by step. Therefore, we looked at tasks given in the International Physics Olympiad~\cite{sciendeolympiade}, an annual physics competition for high school students. This guaranteed that the tasks were suitable, yet challenging, for university students. Of these tasks, we selected four. 
The task texts were adapted in such a way that no picture is necessary for the solution and it was verified that ChatGPT cannot solve the tasks directly and the search engine does not show a page containing the solutions, but both can give hints for obtaining the solution. 
We then designed a pretest containing the selected topics. For this, we created our own questions and took some items from the Energy and Momentum Conceptual Survey (EMCS) v1~\cite{afif2017developing}, 
one item from the appendix of a paper about concepts of force and frictional force~\cite{ sharma2007concepts}, 
and two items from the Rotational Kinematics Concept Inventory~\cite{mashood2012inventory}.
Additionally, we created a question inspired by the sample question on the trade-off between equilibrium and the upper limit on the friction force of the statics concept inventory paper~\cite{steif2005statics}. All pretest questions can be found in the supplementary material.

\subsection{Procedure}
\label{procedure}
\begin{figure}
    \centering
    \includegraphics[width=0.45\textwidth]{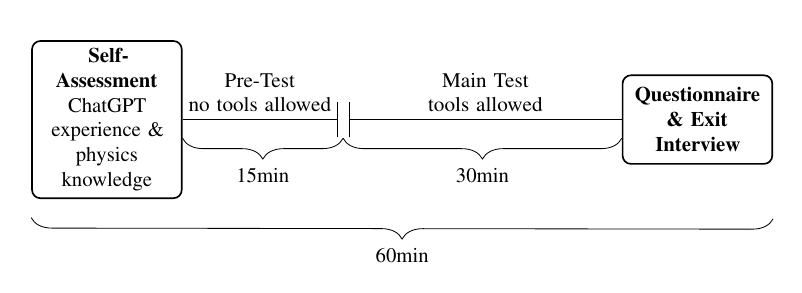}
    \caption{The study procedure timeline in detail. First a self-assessment was conducted, followed by a short pretest and finally the main test, where use of the support tool was allowed. Afterwards followed a questionnaire and depending on the condition an exit interview.}
    \label{fig:procedure}
\end{figure}

The study itself consisted of a survey split into multiple parts, as shown in Figure \ref{fig:procedure}. After providing informed consent and an in-depth explanation of the study procedure, the study started with a self-assessment where participants could rate their physics and ChatGPT knowledge and how often they use ChatGPT. Following that, participants had 15 minutes to solve the 17 multiple-choice pretest questions worth 1 point each (max points = 17). It covered the six physics categories relevant to the main test (see \Cref{question_acquisition}).
After completing the pretest, participants were allowed to use a modified user interface of ChatGPT or a search engine to help them solve the four physics questions (max points = 12) given a time frame of 30 minutes. The written part of the survey ended with a short questionnaire, including the affinity for technology interaction scale~\cite{lezhnina2020multi}, the UMUX-Lite~\cite{lewis2013umux} scale to assess usability, custom questions on perceived accuracy and quality of the tools' answers as well as demographics. Throughout the course of the survey, the order of all questions remained unchanged, ensuring the same experience for all participants. 
For participants attending in person at the university of Kaiserslautern-Landau 
(N=20), we additionally recorded a short (2-5min) exit interview. After the study, participants were reimbursed with the equivalent of \$11 or course credit for a voluntary seminar (N=7). Ethical approval for this study was obtained from the Ethics Committee at the German Research Center for Artificial Intelligence (DFKI).

\subsection{Participants}
For our baseline condition (\baselinecondition), we acquired 13 participants with physics backgrounds from multiple universities, mostly by providing them the option to do the survey online. We further excluded participants who did not fully follow our study instructions, leaving a final number of 12 participants (Age $\samplemean$=23.6\,y, $\samplesd$=2.6\,y, 10m, 2n/a, 3 in person, 9 online). Five studied physics, three electrical engineering, two computer science, one mechanical engineering and one gave no answer.
The students were on average in their eighth semester ($\samplemean$=7.4, $\samplesd$=4.3), scored eight points in the pretest ($\samplemean$=8.2, $\samplesd$=3.9, max=12), had an above-average self-assessed physics knowledge ($\samplemean$=62.8, $\samplesd$=25.1) coupled with below average experience when using ChatGPT ($\samplemean$=40.5, $\samplesd$=33.4)\footnote{Self-assessed physics knowledge and experience using ChatGPT were input on a visual analog scale between 0 and 100.}.

For the second condition of our study (\chatgptcondition), we initially recruited 30 participants from two different universities (RPTU Kaiserslautern-Landau, LMU Munich) 
with a background in physics. They were recruited using mailing lists, posters, and by advertising the study in lectures. However, we had to exclude three students as they did not complete the study, leaving us with 27 participants (Age $\samplemean$=22.6\,y, $\samplesd$=4.0\,y, 25m, 2f, 27 in person, 0 online). 
Participants were, on average, in their sixth semester ($\samplemean$= 5.3, $\samplesd$=3.3). Twelve of them were physics students, three physics and mathematics education students, two techno-physicists, two mathematics students, two electrical engineering students, two electrical engineering and information science students, and one student each from bio- and chemical engineering, computer science, environmental science, and industrial engineering. Participants scored on average nine points in the pretest ($\samplemean$=9.2, $\samplesd$=3.2, max=15). We found no significant difference for the pretest score between the \chatgptcondition{} and \baselinecondition{} conditions. Further, students reported an above-average perceived physics knowledge ($\samplemean$=58.7, $\samplesd$=18.6) and below average experience with ChatGPT ($\samplemean$=42.2, $\samplesd$=24.5).


\subsection{Apparatus}
\label{sec:apparatus}
For the \chatgptcondition{} condition, we used ChatGPT 3.5 turbo with some minor changes on the client side. To allow participants to directly voice their opinions, we implemented a rating scale (good, neutral, bad) to appear with each answer provided by ChatGPT. 
Furthermore, we implemented a download button to be able to save the conversation\footnote{At the time of the study, this feature was not yet available.}. 
For the \baselinecondition{} condition, we set up a website through which participants could use Google while we were able to collect their search queries.

When students participated in person, we further recorded the screen, keystrokes, and mouse movements during the study duration, completing the aforementioned ChatGPT conversation log and the participants' ratings to form a complete interaction protocol and conducted an exit interview.
Additionally, all participants were allowed to use a non-programmable calculator, pen, and paper throughout the study.

\subsection{Measures}\label{sec:metrics}
To allow for a holistic picture of how students interact with \chatgptcondition, we measured student performance through different factors, conducted exit interviews, and analyzed the full student interaction protocols as described in the following section.

\subsubsection{Student Performance}
To evaluate participant performance, first, a grading schema was created by two physics university educators. Using this schema, two other physicists scored the given answers for the four main questions, independently from each other, awarding between zero and three points per question and participant. We evaluated the inter-rater reliability by calculating the average Cohen's Kappa ($\kappa$=0.72) over all main questions, which indicates a substantial reliability~\cite{landis1977measurement}. Zero points represented a completely wrong answer with no correct parts, while three points equated to a completely correct answer. 
Ultimately, both raters reached an agreement in cases where their initial rating differed through discussion. The resulting final scores show student performance in answering the main questions. 
Further, we determined how participants reached their final answers, indicating their problem solving strategy. If the final result of a question was present in the interaction protocol with ChatGPT related to that question, we assigned ``extracted from ChatGPT" as strategy. Otherwise, it was counted as ``own answer". Questions that were not answered were counted as ``none". When it was not evident how the answer was obtained, we assigned ``random guess" as strategy.



\subsubsection{Interaction with the support tools}
We analyzed the interaction of the participants with their respective tool (ChatGPT or search engine). For the \chatgptcondition{} condition, this includes all prompts from participants, respective answers from ChatGPT and associated ratings from participants. Furthermore, for the \baselinecondition{} search queries were analyzed.

\paragraph{Perceived Correctness of ChatGPT Answers}
\label{sec:perceived_correctness}
Having two physicists additionally rate all answers given by ChatGPT for correctness enabled us to compare how students rate answers and their actual correctness. With this information, we were able to estimate two important metrics that demonstrate the students' perception of ChatGPT answers given their actual correctness. A false positive rate (FPR), i.e., ChatGPT answers that students erroneously assumed to be correct (voted positive), and a true positive rate (TPR), i.e., ChatGPT answers that students correctly identified as correct. In our analysis, we focus on these metrics as they highlight how often information from ChatGPT was assumed to be correct. More metrics can be found in the supplementary material.

\paragraph{Interaction Types}
Additionally, we created codes to represent the strategies with which participants created their prompts by categorizing each individual prompt into a coding, comparing and merging them as needed until a consistent representation emerged.


\subsubsection{Custom questions}
As mentioned in \Cref{procedure}, we administered the ATI~\cite{lezhnina2020multi} and UMUX-Lite~\cite{lewis2013umux} as well as two custom questions to inquire about the participants' impression on ChatGPT correctness accuracy and quality.

\subsubsection{Exit Interviews}
The exit interviews were conducted with 20 participants that used \chatgptcondition. In it, we asked the participants five questions with regards to the study. The questions included but were not limited to asking what strategies were used, how the tool was used and how confident participants were in the correctness of their results.

\section{Results}
After explaining how we calculated our measures, we report our results in this section. 

\subsection{Student Performance} 
\label{student_performance}
On average, participants scored $\samplemean$=1.04 points ($\samplesd$=1.43) out of the maximum achievable 12 points in the \chatgptcondition{} condition.
The highest score achieved by a single student was six points.
Three students got more than two points, while twelve students did not score any points at all. 
Most points (nearly 90\%) were achieved in questions Q1 and Q3. We found a large positive correlation between the final score and the pretest score, using Kendall's rank correlation ($\tau$=.37, $p$=.02). No further correlations with respect to the final score were found, in particular for the self-assessed physics knowledge, and study program related demographics.

Analyzing how final answers were obtained, we observed that the most prominent strategy was ``extracted from ChatGPT" being used in 62\% of all cases. Following this, 28\% of participants arrived at their ``own answer", 9\% of questions were not answered (``none") and 1\% made a ``random guess". 

For the \baselinecondition, participants scored $\samplemean$=1.83 points ($\samplesd$=1.27) on average. Four points was the highest amount achieved by two students. Three students achieved more than two points, while one student did not score a single point. Here too, most points (around 95\%) were achieved in questions Q1 and Q3. Using Kendall's rank correlation, we found a statistically significant medium positive correlation ($\tau$=.27, $p$=.03) between the main test score and self-assessed physics knowledge, but none for the pretest score.

Further, we conducted a one-way ANOVA after rank-aligning the data~\cite{wobbrockAlignedRankTransform2011} to investigate whether there are significant differences between our two conditions (\baselinecondition, \chatgptcondition) with regard to the students' performances in the main test. We found that students in the \baselinecondition{} condition performed significantly better ($F(1, 37)$=5.5, $p$=.02, $\eta^2$=.13)\footnote{Effect sizes are given using $\eta^2$ (Partial Eta Squared): small ($>.01$), medium ($>.06$), large ($>.14$).}.

\subsection{Interactions with ChatGPT}
\label{interactions_gpt}
In total, participants working with ChatGPT created 272 prompts, 165 of which were rated (see \Cref{sec:apparatus}). Overall, participants rated 47\% of ChatGPT answers to their prompts as positive, indicating that they deemed them to be correct. 29\% were rated neutral, and 24\% negative, indicating that participants were unsatisfied with them.

\begin{figure}[!ht]
    \centering
    \includegraphics[width=0.45\textwidth]{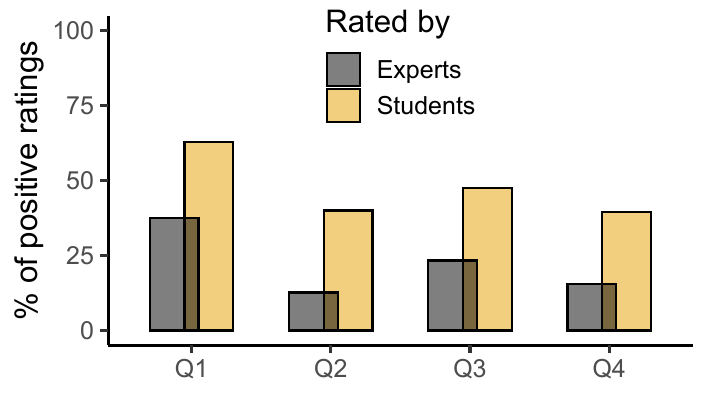}
    \caption{The proportion of positively rated ChatGPT answers to students' prompts visualized for each of the main questions and broken down for students and experts.}
    \label{fig:proportion_s_e}
\end{figure}

Contrarily, our expert physicists only rated approximately 22\% as correct, highlighting a mismatch in expectations. This effect is visible throughout all main questions, as depicted in \Cref{fig:proportion_s_e}. To further analyze intersections in believes of students and experts, we looked at perceived correctness (see \Cref{sec:perceived_correctness}). We obtained a high false-positive rate of 57\%, i.e., over half of all the answers provided by ChatGPT were believed to be correct by participants but rated incorrect by experts. The true-positive rate of 91\%, however, indicates that participants rated most correct answers positive. A complete overview of all metrics can be found in the supplementary material.


\subsubsection{Interaction Types}
\label{interaction_strategy_coding}
We identified four main interaction types based on the reviewed ChatGPT interaction logs from all participants: \emph{copy \& paste}, \emph{preprocessing}, \emph{postprocessing}, and \emph{transformation}. The individual interactions are described in more detail below.

\textbf{Copy \& Paste} is the most prominent interaction type, where participants transferred the physics question directly to ChatGPT without any changes.

\textbf{Preprocessing} is characterized by students trying to reduce the question complexity and using simple priming strategies. They divide a question into multiple parts (P10), ask for formulas (P4), or try to prime the model to improve their results when asking physics questions (P14).

\textbf{Postprocessing} builds on already existing answers given by ChatGPT. The participants try to obtain explanations for parts of a question (P1) or correct mistakes they found in the given answer (P12).
\begin{quoting}
    How do you get the mass of the car and the power of the engine from the question? (P1)
\end{quoting}

\textbf{Transformation} is an interaction type where students used ChatGPT to apply some kind of \emph{transformation} on the data, including translation into another language (P6) and summarizing results~(P3).

\subsubsection{Interaction Strategies}
During the study, we noticed that students built their individual strategies to solve the given physics questions based on these interaction types. For example, a participant might start with priming ChatGPT (\emph{preprocessing}), followed by \emph{copy \& pasting} the question and, ultimately, asking for an explanation of some part of the answer (\emph{post-processing}).

\begin{figure}
    \centering
    \includegraphics[width=0.45\textwidth]{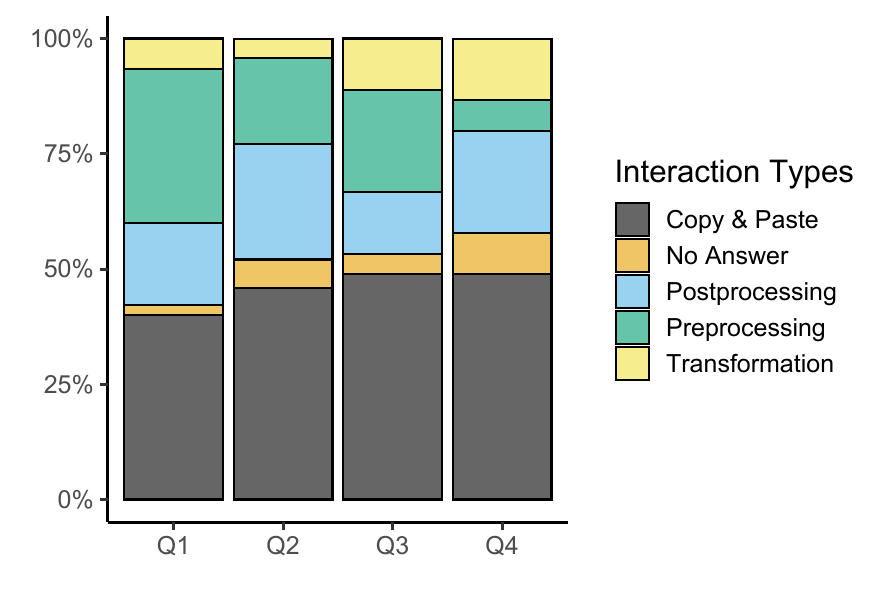}
    \caption{Distribution of interaction types for each question of the main test for the \chatgptcondition{} condition.}
    \label{fig:codings_per_question}
\end{figure}

Overall, \emph{copy \& paste} was the most used interaction strategy, being used 84 times. \emph{Preprocessing}, the next most common strategy, was used 37 times, followed by \emph{post-processing} (36) and \emph{transformation} (16). In \Cref{fig:codings_per_question}, the distribution of used interaction strategies per question is visualized.

\subsection{Interactions with the Search Engine}
\label{interaction_SE}
To be able to compare how participants of both groups interacted with their respective tool, we describe interaction types and strategies when using the search engine here.

\subsubsection{Interaction Types} We divided the interactions done with the search engine into the same four types as the interactions with ChatGPT (\Cref{interactions_gpt}). This allows for easy comparison between the two conditions. There are some minor updates to the interaction types, as some interactions seen when using a search engine were not present when using ChatGPT. \emph{Preprocessing} for the \baselinecondition{} condition mainly consists of asking for formulas and calculations, while \emph{postprocessing} only encompasses asking for explanations. In the \emph{transformation} interaction, the interaction types ``finding answers to related problems" and ``trying to find the initial question using keyword search" were added. There were no changes to the \emph{copy \& paste} interaction.

\subsubsection{Interaction Strategies} The relations between how often different strategies were used changed considerably from the \chatgptcondition{} condition to the \baselinecondition{}. Here, the most used strategy was \emph{preprocessing} with 64 uses, followed by \emph{transformation} with 17 uses, \emph{postprocessing} (8) and \emph{copy \& paste} (3). A corresponding figure of the distribution of interaction types for each question can be found in the supplementary material.


\subsection{Custom Questions}
We calculated the average ATI~\cite{lezhnina2020multi} score of all study participants ($\samplemean$=4.35, $\samplesd$=0.79), which indicates above-average technical affinity. 
Additionally, we used the UMUX-Lite~\cite{lewis2013umux} questionnaire to calculate a parity score for SUS~\cite{brooke1996sus} for the \chatgptcondition{} ($\samplemean$=73.05, $\samplesd$=9.95) and the \baselinecondition{} condition ($\samplemean$=66.23, $\samplesd$=11.62). Both indicate an above-average system usability. Further, participants rated ChatGPT answers for correctness at $\samplemean$=58.0 ($\samplesd$=18.59) and their quality at $\samplemean$=69.26 ($\samplesd$=16.21) on a visual analog scale from 0 to 100. The search engine answer correctness was rated $\samplemean$=59.6 ($\samplesd$=22.8) and its answer quality $\samplemean$=55.5 ($\samplesd$=28.7). We found no significant differences between the two conditions for all custom questions.


\subsection{Exit Interview}
\label{exit_interview}
We recorded the audio of the \chatgptcondition{} exit interviews (59:30 min) and transcribed them verbatim using Whisper~\cite{openaiwhisper2022}, including manual corrections. To analyze the exit interviews, we used the approach by \citet{blandford2016qualitative}. Two researchers coded all interviews separately and merged a final coding tree. From a final discussion, the following themes surfaced: \textsc{strategies}, \textsc{interaction}, and \textsc{reflection} as presented in detail below.

\paragraph{Strategies}
\label{exit_interview_strategies}
While a diverse set of strategies was employed by the participants, most of them mentioned copy \& pasting a question in their exit interview. Different reasons for this were given, such as wanting to see how ChatGPT would deal with the question (P4) or that they did not know how to address the physics question (P7). Other strategies included using ChatGPT like a search engine, e.g., asking for formulas (P1) as it was more convenient. 


Strategies that indicated a higher level of reflection include prior conceptualization of the physics problem asking ChatGPT targeted questions.
\begin{quoting}
    (...) which variables do I know, what kind of information is this, what can I find out with them at hand (...) (P10)
\end{quoting}

Similarly, ChatGPT was used to explore options for possible solutions and approaches. Here, students identified valuable pieces in ChatGPT answers and showed the ability to detect mistakes and inconsistencies in its argumentation.
\begin{quoting}
    (...) there were gaps where it contradicted itself a bit. But it was possible to see what the idea behind it was and whether it made sense. (P4)
\end{quoting}

Though, participants also stated that they had to compromise between time and correctness of their solutions due to the time constraints. While motivated initially, they tried to offload more work to ChatGPT if time was running out (P1).

\paragraph{Interaction} 
When interacting with ChatGPT, participants identified a need to use informed queries. Some tried to achieve this by extracting the most relevant parts of a question from it.
\begin{quoting}
    I have now mostly tried to get the essence out of the questions, so to speak, and then to ask the important ChatGPT, so to speak, not to enter the complete question. (P3)
\end{quoting}

Others found that longer texts worked poorly, implying a need for concrete queries to work around this issue (P20) or requiring participants to dig deeper into an answer given by ChatGPT (P12).

Interestingly, some participants described their interaction/conversation with ChatGPT as human-like, that the answers looked nice and were very well elaborated. However, selected participants feared that this could delude unaware users.
\begin{quoting}
    (...) It also felt very much like I was writing with a person. Very Human (P5)
\end{quoting}
\begin{quoting}
    (...) that the answers are partly quite detailed and one is partly also dazzled by the quantity (P20) 
\end{quoting}


\paragraph{Reflection}
A number of participants were aware that it is important to reflect on the answers given by ChatGPT, rigorously reviewing them for correctness (P20) and identifying mistakes made by ChatGPT.
\begin{quoting}
    But sometimes it has overlooked things or done things wrong or assumed things that were not in the task at all (P1)
\end{quoting}
Especially participants with background knowledge about LLMs were aware of ChatGPT's weaknesses with regard to physics content and knew what to look out for.
\begin{quoting}
    (...) with these mathematical or physical things,(...) you realize that it's a language model and that it's not somehow designed for that. (P18)
\end{quoting}


Contrarily, for most physics questions, participants showed no sign of actively engaging with the exercises, limiting their reflection.
\begin{quoting}
    I copied everything, I typed it in and that was it.~(P14)
\end{quoting}



\section{Discussion}
Our study provides concrete evidence that students heavily relied on answers generated by ChatGPT in their own area of expertise and were not always able to determine their validity. In the following section, we elaborate on these findings and highlight open research questions for the responsible use of LLMs in education.

\paragraph{Overreliance on ChatGPT answers leads to low scores}

Scored student performance \textbf{(RQ1)} was worse than initially expected (\Cref{student_performance}) given our curated selection of exercises. Students using the \chatgptcondition{} condition performed significantly worse compared to students in the \baselinecondition{} condition. Moreover, our study revealed that students in the \chatgptcondition{} condition had difficulties detecting if answers generated by ChatGPT were correct or not, as indicated by their high false positive rate of 57\%.
The unreflected acceptance of presented answers is worrying as it might lead from singular misinformation to general misconception and showcases that there is a definite need to research interactive mechanisms to increase awareness of the uncertainty of LLMs. Contrarily, most search engines are less likely to suffer from this drawback, as presented results are not formulated as answers, a design aspect that could potentially inform the design of future interfaces for LLMs.

\paragraph{Copy \& Paste Is The Most Prominent Strategy for ChatGPT Users}
This overreliance also manifested when analyzing the employed interaction strategies for the two different tools (\Cref{interaction_strategy_coding,interaction_SE}). Nearly all search prompts (96\%) in the \baselinecondition{} condition are systematic, such as extracting keywords or dividing the question. We believe this behavior originates from the inherent nature of the search engine interface. Since it is not structured as a conversation, users focus their prompts on thought-through keywords. In the \chatgptcondition{} condition, 42\% of search prompts are based on \emph{copy \& paste}, highlighting the limited reflection during problem solving when having access to a different interface. While we did observe participants testing out multiple different strategies like priming, reducing the question complexity, or correcting ChatGPT (see \Cref{interaction_strategy_coding}), the majority of questions were answered using the easiest, most convenient option available \textbf{(RQ2)}.

\paragraph{Limitations}
Overall, we expected students to score better given the careful curation of our exercises through physics education researchers (see~\Cref{question_acquisition}). In hindsight, our questions might have been too difficult for a realistic assessment of how students interact with ChatGPT. However, this result also shows that proper training on how to use LLMs such as ChatGPT might be necessary to achieve good results. 
Secondly, the number of total participants that took part in our study was relatively small. To alleviate this, we made the \baselinecondition{} condition of the survey available online as well, allowing us to gather more participants. However, due to the online environment, it is possible that the answer quality was lower compared to in person participants. Though, if that were the case, we can assume that the difference between the two conditions would have been even more prominent. 
\paragraph{A Need for Informed and Moderated Use of LLMs}
Our analysis revealed a need to think about the design of educational systems that use LLMs. We need to moderate interaction with language models such as ChatGPT in a way that students can profit from the vast abilities of such tools while simultaneously reducing the negative impact it can have on the students' learning progress.
We argue that informed use is straightforward to achieve, e.g., through making users aware of the uncertainty of LLMs, especially in the domain of physics. Yet, to really leverage the potential of these models, we must achieve moderated use. In other words, it is a usage that allows students to interact with ChatGPT as a guidance teacher or sparing partner to formulate and explore ideas to solve a physics problem. Such a system should carefully guide students towards the solution, introducing necessary concepts but allowing critical thinking and reflection while still being enjoyable and effective to use.
If we can demonstrate the benefits of moderated LLMs compared to unrestricted LLMs to students in terms of their ability to learn and progress, we can certainly change and evolve the current ways of teaching.


\section{Conclusion}

In this work, we analyzed the impact of having LLMs, such as ChatGPT, as available tools for solving physics exercises on solution correctness and solving strategies. We found that students who used ChatGPT scored significantly lower compared to those using a search engine for the same task. Furthermore, a stark difference in user interaction became visible, where ChatGPT users mainly relied on copy \& pasting questions and answers, while search engine users used more refined strategies such as searching for formulas. This highlights missing reflection and limited critical thinking as two of the main issues when using LLMs in education.
To combat this, we --- first and foremost --- suggest to inform students more adequately of the shortcomings of these models. Though ultimately, we want to converge towards moderated LLMs, specifically designed to support students in a meaningful way by encouraging critical thinking.

\bibliography{sample}

\end{document}